\documentclass[aps,preprint,showpacs,preprintnumbers,amsmath,amssymb]{revtex4}
\usepackage{graphicx}

\def\beq{\begin{equation}}
\def\eeq{\end{equation}}
\def\beqn{\begin{eqnarray}}
\def\eeqn{\end{eqnarray}}

\newcounter{saveeqn}

\def\r {{\bf r}}

\def\d {{\bf d}}

\def\C {{\bf C}}
\def\U {{\bf U}}
\def\D {{\bf D}}
\def\T {{\bf T}}
\def\O {{\bf O}}

\def\bcalH {\mbox{\boldmath $\mathcal H$}}

\def\brho {\mbox{\boldmath $\rho$}}

\def\bphi {\mbox{\boldmath $\phi$}}

\begin{document}

\title{The multi-configurational time-dependent Hartree method for bosons:
           Many-body dynamics of bosonic systems}

\author{Ofir E. Alon\footnote{E-mail: ofir@pci.uni-heidelberg.de},
        Alexej I. Streltsov\footnote{E-mail: alexej@pci.uni-heidelberg.de},
        and Lorenz S. Cederbaum\footnote{E-mail: lorenz.cederbaum@pci.uni-heidelberg.de}}

\affiliation{Theoretische Chemie, Physikalisch-Chemisches Institut, Universit\"at Heidelberg,\\
         Im Neuenheimer Feld 229, D-69120 Heidelberg, Germany}

\begin{abstract}
The evolution of Bose-Einstein condensates is amply 
described by the time-dependent Gross-Pitaevskii mean-field theory
which assumes all bosons to reside in a single time-dependent one-particle state 
throughout the propagation process.
In this work, we go beyond mean-field and develop an 
essentially-exact many-body theory 
for the propagation of the time-dependent Schr\"odinger 
equation of $N$ interacting identical bosons.
In our theory, 
the time-dependent many-boson wavefunction is written as a sum of permanents 
assembled from orthogonal one-particle functions, or orbitals, 
where {\it both} the expansion coefficients {\it and} the permanents (orbitals) themselves 
are {\it time-dependent} and fully determined according to a standard time-dependent variational principle.
By employing either the usual Lagrangian formulation or 
the Dirac-Frenkel variational principle 
we arrive at two sets of coupled equations-of-motion,
one for the orbitals and one for the expansion coefficients. 
The first set comprises of first-order differential equations in time
and non-linear integro-differential equations in position space,
whereas the second set consists of first-order differential equations  
with time-dependent coefficients.
We call our theory multi-configurational time-dependent Hartree 
for bosons, or MCTDHB($M$), where $M$ specifies the number of 
time-dependent orbitals used to construct the permanents.
Numerical implementation of the theory is reported and 
illustrative numerical examples of many-body dynamics of 
trapped Bose-Einstein condensates are provided and discussed.
\end{abstract}

\maketitle

\section{Introduction}

The experimental realizations of Bose-Einstein condensates made of
ultracold alkali-metal atoms 
\cite{Wieman_Cornell_Science_1995,Ketterle_PRL_1995,Ketterle_Nobel_lecture,Cornell_Wieman_Nobel}
have stimulated a modern renaissance as to possible utilization of cold 
trapped atoms and Bose-Einstein condensates,
for instance to quantum information processing \cite{Cirac_Zoller_1995,Brennen_PRL_quan_comp_1999}
and interferometric-based precession measurements \cite{Orzel},
and to the fundamental physics governing trapped interacting bosons 
\cite{Leggett_review,Pethich_book,Stringari_book}.
An ultimate goal of researchers is to be able to design, 
realize, manipulate and detect 
desired quantum states of the many-atom system.

Static and dynamical properties of Bose-Einstein condensates have been extensively
and successfully explored in the community by employing the mean-field 
Gross-Pitaevskii theory \cite{GPO1,GPO2},  
for reviews see \cite{Leggett_review,Pethich_book,Stringari_book} 
and for individual applications to condensates and mixtures 
\cite{R1,R2,R3,R4,R5,R6,R7} and \cite{BB_Ho,BB_Esry,BB_Pu,r21,r22,r23}, respectively.
Gross-Pitaevskii theory is an excellent theory for weakly-interacting 
bosons whenever a single macroscopic one-particle wavefunction is sufficient to describe the reality.
By definition, Gross-Pitaevskii theory cannot describe phenomena such as fragmentation 
of condensates or Mott-insulator phases in optical lattices for which two, 
few or many one-particle functions are occupied.
Recently, we have developed a multi-orbital mean-field approach
to describe static and dynamical properties of fragmented condensates \cite{LA_PLA,OAL_PLA_2007},
thus generalizing the (one-orbital) Gross-Pitaevskii mean-field.
Utilizing the multi-orbital mean-field has enabled us to find new phenomena 
associated with fragmentation, fermionization, quantum phases, 
demixing scenarios and interferences of interacting bosonic systems 
in traps and optical lattices \cite{u1,u2,u3,u4,u5}.

In spite the great successes and popularity of mean-field approaches 
in the many-boson problem of degenerate quantum gases, 
the need and demand for practical many-body theoretical approaches and computational tools
is widely accepted and well documented, see, e.g., 
the review \cite{numerical_review} and references therein.
Nowadays, finite-number condensates can be produced (and reliably reproduced) in experiments.
For instance, see the work of the Raizen group which demonstrated
atom-number squeezing with as little as a few tens of atoms \cite{Raizen_SQ},
and the experiment of the Oberthaler group which demonstrated a single
Josephson junction in a double-well with a few thousands of atoms \cite{Oberthaler_JJ}.
Experiments as those call for delicate description of beyond mean-field, 
finite-particle-number, 
and many-body effects.

The purpose of the present work is to develop an essentially-exact  
and numerically-efficient time-dependent approach for the solution of the 
{\it time-dependent} many-boson Schr\"odinger equation.
Recently, we have developed a general variational theory with complete
self-consistency for studying stationary properties of
condensates in traps on a quantitative many-body level \cite{MCHB_paper}.
The main idea of Ref.~\cite{MCHB_paper} is the computation of 
the best (optimal) stationary many-body wavefunctions expressed
as a linear combination of permanents.
In reminiscence to our approach in the stationary many-boson problem \cite{MCHB_paper}, 
we will here optimize the {\it time-dependent} many-body wavefunctions
according to a time-dependent variational principle.
We term our approach multi-configurational time-dependent Hartree for bosons (MCTDHB). 
The equations-of-motion of MCTDHB have been recently 
posted in their final form in Ref.~\cite{ramp_up_Letter},
where we employed MCTDHB to the popular problem of splitting a 
Bose-Einstein condensate by deforming a single well to a double-well.
Applying MCTDHB to the ramping-up-a-barrier problem we followed the many-boson 
wavefunction throughout the splitting process and identified
the role and impact of many-body excited-states on the splitting process. Among others,
we were able to identify a new 'counter-intuitive' regime where the evolution of 
the condensate when the barrier in ramped-up sufficiently slow is not to the ground-state 
of the double-well which is a fragmented condensate, 
but to a low-lying excited-state which is a coherent condensate \cite{ramp_up_Letter}.
Here we provide the derivation of the MCTDHB theory, 
details of the numerical implementation of MCTDHB,
and complementary illustrative numerical examples.

Evidences that employing time-dependent orbitals 
(permanents) in attacking the time-dependent many-boson Scr\"ordinger equation
beyond Gross-Pitaevskii theory is useful and important 
have already appeared in the literature.
Zoller and co-workers addressed the ramping-up-a-barrier problem
with two time-dependent orbitals of Gaussian shape whose positions and phase change in time \cite{t2}.
More recently, Masiello and Reinhardt have used a
time-dependent multi-configurational ansatz with two orbitals of 
predetermined {\it gerade} and {\it ungerade} symmetries to describe 
interferences in a symmetric double-well set-up \cite{MR_2007}.
In this context, we would like to stress that MCTDHB is fully variational 
and is not restricted to the number of orbitals, 
to a predetermined shape or symmetry of the orbitals, 
and to the geometry, dimensionality and interparticle interactions
of the time-dependent many-boson problem.

It is instructive to 
place the MCTDHB theory developed and presented in this paper 
in the general context of  multi-configurational time-propagation approaches 
of many-particle systems.
The idea to expand and optimize the time-dependent many-body wavefunction
of {\it distinguishable} particles is long known and 
termed multi-configurational time-dependent self-consistent field approach \cite{Makri,Kosloff1,Kosloff2}.
A particular efficient variant of the multi-configurational time-dependent self-consistent field approach
is the multi-configuration time-dependent Hartree (MCTDH) approach
which has been successfully and routinely used for multi-dimensional 
dynamical systems consisting of distinguishable particles \cite{CPL,JCP,CMF,PR}.
 
The MCTDH approach can treat efficiency and accurately static and dynamical properties
of a few-particle systems.
In the latter context, 
we mention that very recently  
static properties of weakly- to strongly-interacting trapped few-boson systems 
have been studied on a quantitative many-body level by MCTDH \cite{ZP1,ZP2}.
Yet, in treating a larger number of identical particles
it is essential to use their statistic properties to
truncate the large amount of redundancies of coefficients
in the distinguishable-particle multi-configurational expansion of the MCTDH wavefunction.
In this case, the challenge was first approached for fermionic systems
where a fermionic version of MCTDH was independently 
developed by several groups \cite{MCTDHF1,MCTDHF2,MCTDHF3}, 
taking explicitly the antisymmetry of the many-fermion wavefunction 
to permutations of any two particles into account.
 
Here we accept the respective challenge for bosons.
This is in particular valuable since  
very-many bosons can reside in only a small number of orbitals.
Alternatively speaking,
by explicitly exploiting bosons' statistics it is possible to successfully and quantitatively 
attack the dynamics of a much large number of bosons with the present MCTDHB theory.
A second importance difference in comparison with MCTDH 
is the nature of interparticle interactions. 
In MCTDHB we employ a general two-body interaction between 
identical bosons whereas MCTDH was designated to treat
nuclear dynamics in which interactions normally involve 
several degrees-of-freedom, or coordinates.

The structure of the paper is as follows.
In section \ref{secII} we develop the working equations of MCTDHB,
highlighting their appealing representation in terms of reduced 
one- and two-body densities and discussing properties of MCTDHB.
In section \ref{secIII} we present details of the numerical 
implementation of MCTDHB.
Illustrative numerical examples are provided in section \ref{secIV},
whereas summary and conclusions are given in section \ref{secV}.
Finally, complementary material is differed to appendices
\ref{A1} and \ref{A2}.

\section{Theory}\label{secII}

The evolution of $N$ interacting structureless bosons is governed 
by the time-dependent Schr\"odinger equation:
\beq\label{MBSE}
 \hat H \Psi = i \frac{\partial \Psi}{\partial t}, \qquad 
 \hat H(\r_1,\r_2,\ldots,\r_N) =  \sum_{j=1}^{N} \hat h(\r_j) +  \sum_{k>j=1}^N W(\r_j-\r_k).
\eeq
Here $\hbar=1$, $\r_j$ is the coordinate of the $j$-th boson, 
$\hat h(\r) = \hat T(\r) + V(\r)$ is the one-body Hamiltonian 
containing kinetic and potential energy terms,
and $W(\r_j-\r_k)$ describes the pairwise interaction between the $j$-th and $k$-th bosons.
In the most general case, the one-body potential $V(\r)$ 
and the two-body interaction $W(\r_j-\r_k)$ and, hence, 
the many-boson Hamiltonian $\hat H$ itself may be time-dependent. 
To avoid cumbersome notation in what follows, 
we do not indicate explicitly
this time-dependence unless it is necessary. 

The time-dependent many-boson Schr\"odinger equation (\ref{MBSE}) cannot be solved exactly (analytically),
except for a few specific cases only, see, e.g., Ref.~\cite{Marvin}.
Hence, theoretical and numerical strategies for attacking this problem are a must.

\subsection{Derivation of the working equations of MCTDHB}

We would like to start by constructing a formally-exact 
representation of the time-dependent many-boson wavefunction $\Psi(t)$
describing the dynamics of $N$ identical structureless bosons.
To this end and following the Introduction part, 
we allow the bosons to occupy permanents made of
time-dependent one-particle functions, or orbitals.
Let us introduce a complete set of time-dependent 
orbitals $\left\{\phi_k(\r,t)\right\}$ which 
are normalized and orthogonal to one another at any time $t$,
\beq\label{orbital_normalization}
 \int \phi^\ast_k(\r,t)\phi_j(\r,t) d\r = \delta_{kj}. 
\eeq
In what follows, it is convenient to work
in second quantization formalism and introduce the set of 
bosonic annihilation operators corresponding to
the orbitals $\left\{\phi_k(\r,t)\right\}$.
This is conveniently done by employing the relation,
\beq\label{annihilation_def}
 b_k(t) = \int \phi_k^\ast(\r,t) \hat{\mathbf \Psi}(\r) d\r,
\eeq
where $\hat{\mathbf \Psi}(\r)=\sum_k b_k(t)\phi_k(\r,t)$ is the usual bosonic 
field operator annihilating a particle at position $\r$. 
The bosonic annihilation and corresponding creation operators 
obey the usual commutation relations 
$b_k(t) b^\dag_j(t) - b^\dag_j(t) b_k(t) = \delta_{kj}$
at any time. 
Using the creation operators $b^\dag_k(t)$ we assemble the permanents,
\beq\label{basic_permanents}
 \left|\vec{n};t\right> =
\frac{1}{\sqrt{n_1!n_2!n_3!\cdots n_M!}} \left(b_1^\dag(t)\right)^{n_1}\left(b_2^\dag(t)\right)^{n_2}
\cdots\left(b_M^\dag(t)\right)^{n_M}\left|vac\right>,
\eeq
where $\vec{n}=(n_1,n_2,n_3,\cdots,n_M)$ represents the occupations of the orbitals
that preserve the total number of particles $n_1+n_2+n_3+\cdots+n_M=N$,
$M$ is a number of the one-particle functions, 
and $\left|vac\right>$ is the vacuum.

In the multi-configuration time-dependent Hartree approach for bosons (MCTDHB)
the {\it ansatz} for the many-body wavefunction $\Psi(t)$ 
is taken as a linear combination of time-dependent permanents (\ref{basic_permanents}) \cite{ramp_up_Letter},
\beq\label{MCTDHB_Psi}
\left|\Psi(t)\right> = 
\sum_{\vec{n}}C_{\vec{n}}(t)\left|\vec{n};t\right>,
\eeq
where the summation runs over all possible configurations whose 
occupations $\vec{n}$ preserve the total number of bosons $N$.
Of course, in the limit $M \to \infty$ the set of permanents $\{\left|\vec{n};t\right>\}$
spans the complete $N$-boson Hilbert space and thus expansion (\ref{MCTDHB_Psi}) is exact.
So where is the advantage of utilizing an expansion with time-dependent permanents?
In practice, we have of course to limit the size of the Hilbert space exploited in computations.
Now, by allowing also the permanents to be time-dependent we 
can use much shorter expansions than if only
the expansion coefficients are taken to be time-dependent,
thus leading to a significant computational advantage.
We would like to highlight that in representation (\ref{MCTDHB_Psi})
both the expansion coefficients $\{C_{\vec{n}}(t)\}$ and 
orbitals $\{\phi_k(\r,t)\}$ comprising the permanents $\left|\vec{n};t\right>$
are independent parameters. 
To solve for the time-dependent wavefunction $\Psi(t)$  
means to determine the evolution of the coefficients $\{C_{\vec{n}}(t)\}$ 
and orbitals $\{\phi_k(\r,t)\}$ in time.

To derive the equations-of motion governing the evolution 
of $\{C_{\vec{n}}(t)\}$ and $\{\phi_k(\r,t)\}$
we need to employ a time-dependent variational principle.
Two such variational principles are utilized here, 
both leading to the same result, of course, 
but highlighting complementary aspects when treating the time-dependent many-boson problem. 
Specifically, in the bulk of the paper below we employ the usual Lagrangian formulation \cite{LF1,LF2},
whereas how to derive the MCTDHB equations-of-motion starting 
from the Dirac-Frenkel variational principle \cite{DF1,DF2} is deferred to appendix \ref{A2}.
The main difference between the two (equivalent) formulations employed here is that,
in the Lagrangian formulation we are to take first expectation values and only subsequently 
perform the variation which somewhat simplifies the algebra, 
whereas in the Dirac-Frenkel formulation the situation reverses, i.e., 
variation of $\Psi(t)$ precedes the computation of matrix elements.

In the framework of the Lagrangian formulation \cite{LF1,LF2},
we substitute the many-body {\it ansatz} (\ref{MCTDHB_Psi}) for $\Psi(t)$
into the functional action of the time-dependent Schr\"odinger equation which reads:
\beq\label{action_functional}
 S\left[\{C_{\vec{n}}(t)\},\{\phi_k(\r,t)\}\right]
 = \int dt \left\{ \left<\Psi\left|\hat H - i\frac{\partial}{\partial t} \right|\Psi\right>
 - \sum_{k,j}^{M} \mu_{kj}(t)\left[\left<\phi_k|\phi_j\right> - \delta_{kj} \right]\right\}. 
\eeq
The time-dependent Lagrange multipliers $\mu_{kj}(t)$ are introduced to ensure that
the time-dependent orbitals $\phi_k(\r,t)$ remain orthonormal throughout the propagation, 
see (\ref{orbital_normalization}) and appendix \ref{A2}.
The next step is to require stationarity of the 
action with respect to its arguments $\{C_{\vec{n}}(t)\}$ and $\{\phi_k(\r,t)\}$.
This variation is performed below separately for
the coefficients and for the orbitals,
recalling that they are independent variational parameters.

\subsubsection{Variation with respect to the expansion coefficients $\{C_{\vec{n}}(t)\}$}\label{sub_small1}

To take the variation of the functional action (\ref{action_functional})
with respect to the expansion coefficients $\{C_{\vec{n}}(t)\}$
we first express the expectation value in the action in a form which 
explicitly depends on the expansion coefficients,
\beq\label{expectation1_C}
\left<\Psi\left|\hat H - i\frac{\partial}{\partial t} \right|\Psi\right> = 
\sum_{\vec{n}} C^\ast_{\vec{n}} 
\left[\sum_{\vec{n}'}\left<\vec{n};t\left|\hat H - 
i\frac{\partial}{\partial t}\right|\vec{n}';t\right> C_{\vec{n}'} - 
i\frac{\partial C_{\vec{n}}}{\partial t}\right].
\eeq
It is now straightforward to perform this variation which gives,
\beq\label{variation1_C}
 \frac{\partial S\left[\{C_{\vec{n}}(t)\},\{\phi_k(\r,t)\}\right]}{\partial C^\ast_{\vec{n}}(t)} = 0
\qquad \Longrightarrow \qquad 
\sum_{\vec{n}'}\left<\vec{n};t\left|\hat H - 
i\frac{\partial}{\partial t}\right|\vec{n}';t\right> C_{\vec{n}'} = 
i\frac{\partial C_{\vec{n}}}{\partial t}.
\eeq
Defining the time-dependent matrix $\bcalH(t)$ the elements of which are
\beq\label{Floquet_permanent_matrix_elemets}
{\mathcal H}_{\vec{n}\vec{n}'}(t) =
 \left<\vec{n};t\left|\hat H - i\frac{\partial}{\partial t}\right|\vec{n}';t\right>,
\eeq
and collecting the expansion coefficients $C_{\vec{n}}(t)$ in the column vector $\C(t)$,
Eq.~(\ref{variation1_C}) can be written in a compact matrix form as:
\beq\label{variation1_C_matrix}
 \bcalH(t) \C(t) = i\frac{\partial \C(t)}{\partial t}.
\eeq
Let us discuss the properties of Eq.~(\ref{variation1_C_matrix}).
The equation-of-motion (\ref{variation1_C_matrix}) is a first-order differential equation in time.
The matrix $\bcalH(t)$ multiplying the vector $\C(t)$ on the 
left-hand-side is time-dependent,
since it is evaluated with time-dependent permanents 
$\left|\vec{n};t\right>$ and $\left|\vec{n}';t\right>$, 
comprised themselves of time-dependent orbitals. 

In the course of evolution, the many-body wavefunction $\Psi(t)$ should,
of course, remain normalized since for self-adjoined Hamiltonians $\hat H$ the evolution is unitary.
$\left|\Psi(t)\right>=\sum_{\vec{n}}C_{\vec{n}}(t)\left|\vec{n};t\right>$ 
would remain normalized if the vector of coefficients $\C(t)$ remains normalized in time 
throughout propagation via Eq.~(\ref{variation1_C_matrix}) 
(the orbitals comprising the permanents remain orthonormal to one another
by virtue of the Lagrange multipliers, see subsection \ref{orbital_variation} for more details).
For this, the matrix $\bcalH(t)$ should be hermitian.

Below, we first evaluate the matrix $\bcalH(t)$ explicitly and then show that it is always hermitian. 
Derivation of the rules to evaluate matrix elements 
of one- and two-body operators between two general permanents can be found in \cite{MCHB_paper}. 
Here the matrix elements with respect to two general time-dependent permanents 
$\left|\vec{n};t\right>$ and $\left|\vec{n}';t\right>$ are displayed in their final form.
Noticing that in second quantization the time-derivative can be treated as a one-body operator,
\beq\label{time_SQ}
 i\frac{\partial}{\partial t} =  \sum_{k,q} b_k^\dag b_q 
   \left(i\frac{\partial}{\partial t}\right)_{kq},
\eeq
the non-vanishing matrix elements of the one-body operator $\hat h -i\frac{\partial}{\partial t}$ are given by
\beqn\label{matrix_elements_1b}
& & \left<\vec{n};t\left|\hat h -i\frac{\partial}{\partial t}\right|\vec{n};t\right> =
\sum_{l=1}^{M} n_l 
\left[h_{ll} - \left(i\frac{\partial}{\partial t}\right)_{ll}\right], 
\nonumber \\
& & \left<\vec{n}_q^k;t\left|\hat h -i\frac{\partial}{\partial t}\right|\vec{n};t\right> =
 \sqrt{(n_k+1)n_q} \left[h_{kq} - \left(i\frac{\partial}{\partial t}\right)_{kq}\right], \ k\ne q, \
\eeqn
where the permanent denoted by $\left|\vec{n}_q^k;t\right>$ differs from $\left|\vec{n};t\right>$
by an excitation of one boson from the $q$-th to the $k$-th orbital.  
The corresponding matrix elements of the two-body operator 
$\hat W$ between two permanents $\left|\vec{n};t\right>$ and $\left|\vec{n}';t\right>$ 
are collected for convenience in appendix \ref{A1}.
The time-dependent matrix elements of the one- and two-body operators 
with respect to orbitals needed in this work are given by:
\beqn\label{matrix_elements}
 h_{kq} &=& \int \phi_k^\ast(\r,t) \hat h(\r) \phi_q(\r,t) d\r, \nonumber \\
 \left(i\frac{\partial}{\partial t}\right)_{kq} &=& 
i\int \phi_k^\ast(\r,t) \frac{\partial\phi_q(\r,t)}{\partial t} d\r, \nonumber \\
W_{ksql} &=& \int \!\! \int \phi_k^\ast(\r,t) \phi_s^\ast(\r',t) \hat W(\r-\r')  
 \phi_q(\r,t) \phi_l(\r',t) d\r d\r', \nonumber \\
 W_{ks[ql]} &=& W_{ksql} + W_{kslq}. 
\eeqn
Note the plus sign in $W_{ks[ql]}$ which is due to bosons' statistics.

With the elements of $\bcalH(t)$ explicitly expressed,
we can return now to the question of its hermiticity.
Clearly, the contribution originating from the matrix elements of the 
Hamiltonian, $h_{kq}$ and $W_{ksql}$, is hermitian for any set of 
of functions $\{\phi_k(\r,t)\}$ belonging to the domain of the Hamiltonian.
To show that the matrix representation 
of $i\frac{\partial}{\partial t}$ is also hermitian
we make use of the fact that the orbitals $\{\phi_k(\r,t)\}$
are kept normalized and orthogonal to one another at any time. 
Taking the time derivative of Eq.~(\ref{orbital_normalization}) we readily get
\beq
 i\frac{\partial}{\partial t} \left[\int \phi^\ast_k(\r,t)\phi_q(\r,t) d\r\right] = 0 
\quad \Longrightarrow \quad \left(i\frac{\partial}{\partial t}\right)_{kq} =
 \left(i\frac{\partial}{\partial t}\right)^\ast_{qk}.
\eeq
Summarizing, we have shown that the matrix ${\bcalH}(t)$
is hermitian for a general set of orbitals $\{\phi_k(\r,t)\}$
that are kept normalized and orthogonal to one another at any time.
Consequently, the evolution of the expansion coefficients $\{C_{\vec{n}}(t)\}$ is always unitary,
namely, that an initially normalized expansion coefficient vector $\C(t)$ propagated 
by Eq.~(\ref{variation1_C_matrix}) remains normalized at any time. 
This result, together with Eq.~(\ref{orbital_normalization}),
guarantees that an initially normalized many-body state $\Psi(0)$ 
remains normalized at any time.

\subsubsection{Variation with respect to the orbitals $\{\phi_k(\r,t)\}$}\label{orbital_variation}

In the present subsection we derive the equations-of-motion governing the 
evolution of the orbitals $\{\phi_k(\r,t)\}$.
Now, it is helpful to express the expectation value of 
$\hat H -i\frac{\partial}{\partial t}$ appearing in the 
functional action (\ref{action_functional})
in a form which allows one for a direct functional differentiation 
with respect to $\phi_k(\r,t)$.

To this end, we write the operator $\hat H -i\frac{\partial}{\partial t}$ in second quantization form:
\beq\label{MB_floquet}
 \hat H -i\frac{\partial}{\partial t} =
 \sum_{k,q} b_k^\dag b_q 
\left[h_{kq} - \left(i\frac{\partial}{\partial t}\right)_{kq}\right]
 + \frac{1}{2}\sum_{k,s,q,l} b_k^\dag b_s^\dag b_q b_l W_{ksql}, 
\eeq
where the matrix elements of the one- and two-body operators are
given in Eq.~(\ref{matrix_elements}).
In calculating the expectation value of (\ref{MB_floquet}) with respect to
the many-boson wavefunction $\Psi(t)$ it is gratifying to make use of
the one- and two-body density matrices $\rho_{kq}(t)=\left<\Psi\left|b_k^\dag b_q\right|\Psi\right>$
and $\rho_{ksql}(t)=\left<\Psi\left|b_k^\dag b_s^\dag b_q b_l\right|\Psi\right>$, respectively.
Given the (normalized) wavefunction $\Psi(t)$, the one-body density matrix is given by
\beqn\label{DNS1}
 \rho(\r_1|\r'_1;t) &=&  N\int \Psi^\ast(\r'_1,\r_2,\ldots,\r_N;t) 
 \Psi(\r_1,\r_2,\ldots,\r_N;t) d\r_2 d\r_3 \cdots d\r_N = \nonumber \\
&=& \left<\Psi(t)\left|\hat{\mathbf \Psi}^\dag(\r'_1)\hat{\mathbf \Psi}(\r_1)\right|\Psi(t)\right> =
 \sum^M_{k,q=1} \rho_{kq}(t) \phi^\ast_k(\r'_1,t)\phi_q(\r_1,t), \
\eeqn
where the matrix elements of the density explicitly read 
\beqn\label{DNS1_matrix_elements}
& & \rho_{kk}(t) = 
\sum_{\vec n} C_{\vec n}^\ast C_{\vec n} n_k, \nonumber \\
& & \rho_{kq}(t) = \sum_{\vec n} C_{\vec n}^\ast C_{\vec n_k^q} \sqrt{n_k(n_q+1)}, \ k\ne q. \
\eeqn
It is convenient to collect these matrix elements as $\brho(t)=\left\{\rho_{kq}(t)\right\}$.
Similarly, the two-body density matrix associated with $\Psi(t)$ is given by
\beqn\label{DNS2}
& & \!\!\! \rho(\r_1,\r_2|\r'_1,\r'_2;t) = 
N(N-1)\int\Psi^\ast(\r'_1,\r'_2,\r_3,\ldots,\r_N;t) \Psi(\r_1,\r_2,\r_3,\ldots,\r_N;t)
     d\r_3 \cdots d\r_N = \nonumber \\
& & \!\!\! =\left<\Psi(t)\left|\hat{\mathbf \Psi}^\dag(\r'_1)\hat{\mathbf \Psi}^\dag(\r'_2)
\hat{\mathbf \Psi}(\r_1)\hat{\mathbf \Psi}(\r_2)\right|\Psi(t)\right> =
 \sum^M_{k,s,q,l=1} \rho_{ksql}(t) 
\phi^\ast_k(\r'_1,t) \phi^\ast_s(\r'_2,t) \phi_q(\r_1,t) \phi_l(\r_2,t), \nonumber \\ 
\eeqn
where the matrix elements $\rho_{ksql}$ are collected for convenience in appendix \ref{A1}.

Combining the above ingredients we obtain for the expectation value
of the operator $\hat H -i\frac{\partial}{\partial t}$ appearing in the
factional action (\ref{action_functional}),
\beq\label{expectation2_phi}
 \left<\Psi\left|\hat H - i\frac{\partial}{\partial t} \right|\Psi\right> = 
 \sum_{k,q=1}^M \rho_{kq} \left[ h_{kq} - \left(i\frac{\partial}{\partial t}\right)_{kq} \right]
 + \frac{1}{2}\sum_{k,s,q,l=1}^M \rho_{ksql} W_{ksql} - i\sum_{\vec{n}} C^\ast_{\vec{n}} 
\frac{\partial C_{\vec{n}}}{\partial t}. 
\eeq
Note the appealing appearance of representation (\ref{expectation2_phi}) 
in which the only explicit dependence on the orbitals $\{\phi_k(\r,t)\}$ 
is grouped into the matrix elements 
$h_{kq},\,\left(i\frac{\partial}{\partial t}\right)_{kq}$ and $W_{ksql}$
of the one- and two-body operators $\hat h - i\frac{\partial}{\partial t}$ and $\hat W$, respectively,
whereas the elements $\rho_{kq}$ and $\rho_{ksql}$ of the density matrices do not depend 
explicitly on the orbitals.
Consequently, it is now straightforward to perform variation of the 
functional action (\ref{action_functional}) with respect to the orbitals $\{\phi_k(\r,t)\}$
which gives the following set of coupled equations-of-motion, $k=1,\ldots,M$, 
\beq\label{MCTDHB_equations_general} 
\frac{\delta S\left[\{C_{\vec{n}}(t)\},\{\phi_k(\r,t)\}\right]}{\delta\phi^\ast_k(\r,t)}=0 
\ \ \Longrightarrow \ \ \sum^M_{q=1}\left[\rho_{kq} \left(\hat h - i\frac{\partial}{\partial t} \right) +
\sum^M_{s,l=1}\rho_{ksql} \hat{W}_{sl}\right]\left|\phi_q\right>
= \sum^M_{j=1}\mu_{kj}\left|\phi_j\right>,
\eeq
where
\beq\label{one_body_pot}
  \hat W_{sl}(\r,t) = \int \phi_s^\ast(\r',t) W(\r-\r') \phi_l(\r',t) d\r'
\eeq
are {\it local} time-dependent potentials.

To proceed, we can use the constraints (\ref{orbital_normalization}) 
that the $\phi_k(\r,t)$ are functions orthonormal to one another 
in order to eliminate the Lagrange multipliers $\mu_{kj}(t)$ from Eq.~(\ref{MCTDHB_equations_general}).
By taking the scalar product of $\left<\phi_j\right|$ 
with (\ref{MCTDHB_equations_general}), 
the resulting $\mu_{kj}(t)$ take on the form,
\beq\label{MCTDHB_mu}
 \mu_{kj}(t) =  
 \sum^M_{q=1}\left\{\rho_{kq} \left[ h_{jq} - \left(i\frac{\partial}{\partial t}\right)_{jq} \right] +
 \sum^M_{s,l=1}\rho_{ksql} W_{jsql}\right\}. 
\eeq 
Substituting Eq.~(\ref{MCTDHB_mu}) into (\ref{MCTDHB_equations_general}), 
employing the identities 
\beqn\label{MCTDHB_equ_0}
& &  \sum^M_{q=1}\left[\rho_{kq} \left(\hat h - i\frac{\partial}{\partial t} \right) +
\sum^M_{s,l=1}\rho_{ksql} \hat{W}_{sl}\right]\left|\phi_q\right>
 - \sum^M_{j=1}\mu_{kj'}\left|\phi_j'\right> = \nonumber \\
& & \left(1-\sum_{j'=1}^{M}\left|\phi_{j'}\left>\right<\phi_{j'}\right|\right)
 \sum^M_{q=1}\left[\rho_{kq} \left(\hat h - i\frac{\partial}{\partial t} \right) +
\sum^M_{s,l=1}\rho_{ksql} \hat{W}_{sl}\right]\left|\phi_q\right>, \quad k=1,\ldots,M,
\eeqn
and multiplying from the left with the inverse
of the one-body density and summing over $\sum_k \left\{\brho(t)\right\}^{-1}_{jk}$, 
we arrive immediately at the following form 
of the equations-of-motion of the orbitals $\phi_j(\r,t)$, $j=1,\ldots,M$:
\beqn\label{MCTDHB0_equ}
& & 
  \hat {\mathbf P} i\left|\dot\phi_j\right> = 
  \hat {\mathbf P} \left[\hat h \left|\phi_j\right>  + \sum^M_{k,s,q,l=1} 
  \left\{\brho(t)\right\}^{-1}_{jk} \rho_{ksql} \hat{W}_{sl} \left|\phi_q\right> \right], \nonumber \\
& & \qquad \hat {\mathbf P} = 1-\sum_{j'=1}^{M}\left|\phi_{j'}\left>\right<\phi_{j'}\right|. \
\eeqn
Here and hereafter we use the shorthand notation $\dot \phi_j \equiv \frac{\partial\phi_j}{\partial t}$.
Examining Eq.~(\ref{MCTDHB0_equ}) we see that
eliminating the Lagrange multipliers $\mu_{kj}(t)$
has emerged as a projection operator $\hat {\mathbf P}$ 
onto the subspace orthogonal to that spanned by the $\{\phi_k(\r,t)\}$. 
This projection appears both on the left- and right-hand-sides of (\ref{MCTDHB0_equ}),
making (\ref{MCTDHB0_equ}) a cumbersome coupled system of integro-differential non-linear equations.

Fortunately, due to the invariance properties of the ansatz (\ref{MCTDHB_Psi}), 
see discussion in subsection \ref{MCTDHB_properties} below, 
we can always perform a unitary transformation without introducing further constraints
on the orbitals such that \cite{CPL,JCP} 
\beq\label{MCTDHB_const}
\left<\phi_k|\dot\phi_q\right> = 0, \ \ k,q=1,\ldots,M,
\eeq
are satisfied at any time.
Obviously, if conditions (\ref{MCTDHB_const}) are satisfied at any time,
the orthogonality constraints (\ref{orbital_normalization}) are also satisfied.
This representation simplifies considerably the equations-of-motion (\ref{MCTDHB0_equ}) 
for the orbitals $\phi_j(\r,t)$, $j=1,\ldots,M$:
\beqn\label{MCTDHB1_equ}
& & 
  i\left|\dot\phi_j\right> = \hat {\mathbf P} \left[\hat h \left|\phi_j\right>  + \sum^M_{k,s,q,l=1} 
  \left\{\brho(t)\right\}^{-1}_{jk} \rho_{ksql} \hat{W}_{sl} \left|\phi_q\right> \right], \nonumber \\ 
& &  
\qquad \hat {\mathbf P} = 1-\sum_{j'=1}^{M}\left|\phi_{j'}\left>\right<\phi_{j'}\right|. \
\eeqn
The ${\mathbf P}$ remaining on the right-hand-side of Eq.~(\ref{MCTDHB1_equ}) makes it clear that  
conditions (\ref{MCTDHB_const}) are indeed satisfied at any time throughout the
propagation of $\{\phi_k(\r,t)\}$. 
In practice, the meaning of these conditions is that the temporal changes of the $\{\phi_k(\r,t)\}$ 
are always orthogonal to the $\{\phi_k(\r,t)\}$ themselves.
This property also utilized in MCTDH \cite{CPL,JCP} 
generally makes the time propagation of Eq. (\ref{MCTDHB1_equ}) robust and stable 
and can thus be exploited to maintain accurate propagation results at lower computational costs.
Additionally with conditions (\ref{MCTDHB_const}), 
Eq.~(\ref{variation1_C_matrix}) now reads:
\beq\label{variation2_C_matrix}
 {\mathbf H}(t)\C(t) = i\frac{\partial \C(t)}{\partial t},
\eeq   
where ${\mathbf H}(t)$ is the Hamiltonian matrix the elements of which are 
$H_{\vec{n}\vec{n}'}(t) = \left<\vec{n};t\left|\hat H\right|\vec{n}';t\right>$.
The coupled equation sets (\ref{MCTDHB1_equ}) for the orbitals $\{\phi_j(\r,t)\}$ 
and (\ref{variation2_C_matrix}) for the expansion coefficients $\{C_{\vec{n}}(t)\}$ 
are at the heart of the multi-configurational time-dependent Hartree theory for bosons 
-- a formally-exact and practical representation of the time-dependent 
many-boson Schr\"odinger equation (\ref{MBSE}).
As mentioned above, we term our theory in short MCTDHB($M$), 
where $M$ stands for the number of 
time-dependent orbitals comprising the permanents.
Of course, in the limit $M \to \infty$ the set of permanents $\{\left|\vec{n};t\right>\}$
spans the complete $N$-boson Hilbert space and thus $\Psi(t)$ is exact.
In fact, inspecting Eq.~(\ref{MCTDHB1_equ}) and the projector ${\mathbf P}$ 
therein tells us that in this limit the orbitals $\{\phi_k(\r,t)\}$ become time-independent,
similarly to the situation in the general MCTDH theory \cite{CPL,JCP}.
So where is the advantage of utilizing time-dependent permanents?
In practice, we have of course to limit the size of the Hilbert space used in computations.
By allowing the permanents to be time-dependent, 
variationally-optimized
quantities we can use much shorter expansions than if 
permanents comprised of fixed-shape orbitals are employed,
thus leading to a significant computational advantage.

\subsection{Properties of MCTDHB and its working equations}\label{MCTDHB_properties}

As mentioned in the Introduction,
it is gratifying to note that the present many-body propagation 
theory -- MCTDHB -- adapts to identical bosons the multi-configurational
time-dependent Hartree (MCTDH) approach routinely used for multi-dimensional dynamical 
systems consisting of distinguishable particles \cite{CPL,JCP,CMF,PR}. 
By explicitly exploiting bosons' statistics 
and the fact that only a two-body interaction is needed
it is possible to successfully and 
quantitatively attack the dynamics of a large number of bosons with MCTDHB.
Hence, many of the properties of MCTDHB are inherited from MCTDH.
In this section we concentrate on and expand those
properties that are required for our needs.

The ansatz (\ref{MCTDHB_Psi}) for the many-boson wavefunction includes
all possible permanents $\left|\vec{n};t\right>$ assembled when distributing $N$ bosons 
over $M$ (time-dependent) orbitals $\phi_k(t)$. 
Since the above defines complete Hilbert subspaces,
$\left|\Psi(t)\right>$ possesses invariance properties, i.e., it does
not have a unique representation in terms of the orbitals and coefficients.
Specifically, we can introduce an $M\times M$ unitary matrix $\U(t)=\left\{U_{kq}(t)\right\}$,
define a new set of orthonormal orbitals $\widetilde{\phi}_q(t)=\sum_k U_{kq}(t)\phi_k(t)$,
and assemble all possible permanents $\left|\widetilde{\vec{n};t}\right>$ with them.
Correspondingly, the vector of coefficients is transformed and (\ref{MCTDHB_Psi}) can be 
rewritten to express this invariance,
\beq\label{invariance}
 \left|\Psi(t)\right> = 
\sum_{\vec{n}}C_{\vec{n}}(t)\left|\vec{n};t\right> =
\sum_{\vec{n}}\widetilde{C}_{\vec{n}}(t)\left|\widetilde{\vec{n};t}\right>.
\eeq 
The size of the Hilbert subspace remains the same, of course.
To express that we use the same vector of enumeration $\vec{n}$. 

The invariance of $\left|\Psi(t)\right>$ to unitary transformations of the orbitals
has been used in subsection \ref{orbital_variation} to
simplify the form (\ref{MCTDHB0_equ}) and (\ref{variation1_C_matrix}) of the equations-of-motion
to the respective form (\ref{MCTDHB1_equ}) and (\ref{variation2_C_matrix}) of MCTDHB
which is amenable to numerical implementation.
This has been achieved by employing the differential conditions (\ref{MCTDHB_const}).
We stress that an invariance like (\ref{invariance})
has been used by the developers of MCTDH in order to choose the differential form (\ref{MCTDHB_const})
to simplify the MCTDH equations-of-motion \cite{CPL,JCP}.
Although known from their original MCTDH papers for distinguishable particles \cite{CPL,JCP},
it is deductive to stipulate an explicit proof that 
employing condition (\ref{MCTDHB_const}) throughout the time evolution
does not introduce further constraints or approximations on the variational treatment of the 
time-dependent many-boson (many-body) Schr\"odinger equation.
To this end, we show by direct construction that there is a unitary matrix $\U(t)$
that carries a {\it general} set of time-dependent orthogonal orbitals
to a {\it specific} set of time-dependent orthogonal orbitals satisfying Eq.~(\ref{MCTDHB_const}). 

To obtain an equation-of-motion for $\U(t)$,
we start from the hermitian matrix $D_{kq}(t)=i\left<\phi_k(t)|\dot\phi_q(t)\right>$,
whose elements are not necessarily zero numbers,
and wish to compute the unitary transformation 
$\widetilde{\phi}_q(t)=\sum_{k=1}^M U_{kq}(t)\phi_k(t)$
which guarantees the relations $\widetilde{D}_{kq}(t)=
i\left<\widetilde{\phi}_k(t)|\dot{\widetilde{\phi}}_q(t)\right>=0$. 
For this, we substitute the time-derivative
$\dot{\widetilde{\phi}}_q(t)=\sum_{k=1}^M\left\{\dot U_{kq}(t)\phi_k(t) + U_{kq}(t)\dot\phi_k(t)\right\}$
into the equation $\widetilde{D}_{kq}(t)=0$  
and solve for the matrix $\U(t)$ which ensures these conditions at all times.
Making explicit use of the assumption that $\U(t)$ is unitary we 
immediately get and {\it symbolically} integrate,
\beq\label{U_explicit_solution}
 i \dot U_{sq}(t)= - \sum_{k=1}^M D_{sk}(t) U_{kq}(t) \qquad \Longrightarrow \qquad
  \U(t) = e^{+i \int^t \D(t') dt'} \U(0).
\eeq
Obviously, since $\D(t)$ is an hermitian matrix, 
$\U(t)$ remains unitary at all times if and only if the initial condition $\U(0)$ is an unitary matrix.
We show now by direct construction that $\U(0)\equiv\lim_{\tau\to 0}\U(\tau)$ is unitary and unique.
For this we diagonalize the hermitian matrix $\D(t)$
with the help of the unitary matrix $\T(t)$,
\beq\label{diagonalize}
 \T^\dag(t) \D(t) \T(t) = \d(t),
\eeq
where $\d(t)$ is the diagonal matrix of the eigenvalues of $\D(t)$.
From this it is not hard to see that in the limes $\tau \to 0$,
\beq\label{U0_solution}
 \U(\tau) = \T(0) e^{+i \tau \d(0)},\ \ 
\widetilde{\phi}_q(\tau) = \sum_{k=1}^M U_{kq}(\tau)\phi_k(\tau) \ \ 
 \ \ \Longrightarrow \ \ \ \ \widetilde{D}_{kq}(\tau)=0,
\eeq
which concludes our proof.

Thus, for any set of initial conditions $\{\phi_k(0)\}$ 
we can choose the constraint (\ref{MCTDHB_const}), i.e., $D_{kq}(t)=0$,
and safely work with equations-of-motion 
(\ref{MCTDHB1_equ}) and (\ref{variation2_C_matrix}) of MCTDHB
without loss of generality.
In fact, as has been shown in the original MCTDH work \cite{JCP},
we can use a larger class of constraints, namely 
$D_{kq}(t)=\left<\phi_k\left|\hat g\right|\phi_q\right>$,
where $\hat g$ is a self-adjoint operator in the subspace of orbitals $\left\{\phi_k(\r,t)\right\}$.
The proof that such a choice 
only amounts to a time-dependent unitary transformation of the orbitals and
leads to no further restrictions follows exactly the same lines as above. 
Here we just write the final result for the equations-of-motion which take the form, $j=1,\ldots,M$:
\beqn\label{MCTDHB2_equ}
& & 
  i\left|\dot\phi_j\right> = \hat g \left|\phi_j\right> + 
  \hat {\mathbf P} \left[\left(\hat h - \hat g\right) \left|\phi_j\right>  + \sum^M_{k,s,q,l=1} 
  \left\{\brho(t)\right\}^{-1}_{jk} \rho_{ksql} \hat{W}_{sl} \left|\phi_q\right> \right], \nonumber \\ 
& & {\bcalH}(t)\C(t) = i\frac{\partial \C(t)}{\partial t}, \
\eeqn 
where now ${\bcalH}(t)$ is the time-dependent matrix with elements
${\mathcal H}_{\vec{n}\vec{n}'}(t) = \left<\vec{n};t\left|\hat H - \hat g\right|\vec{n}';t\right>$.

The next property we would like to address is that of the MCTDHB energy, 
$\left<\Psi(t)\left|\hat H \right|\Psi(t)\right>$. 
For time-independent Hamiltonians, as has been shown for MCTDH itself \cite{CPL,JCP},
the energy is constant in time as it should be
for time-evolution with such Hamiltonians. 
For a time-dependent Hamiltonian $\hat H(t)$, 
it is not difficult to show by direct differentiation of the MCTDHB energy
and utilizing without loss of generality relations (\ref{MCTDHB_const}) that, 
\beq\label{e_of_t}
 \frac{d}{dt}\left<\Psi(t)\left|\hat H \right|\Psi(t)\right> =
 \left<\Psi(t)\left|\frac{\partial \hat H}{\partial t}\right|\Psi(t)\right>.
\eeq 
This appealing relation can be used in numerical calculations to monitor 
the degree of accuracy of integration.

Finally, there are two points worth mentioning.
First, we can also propagate the MCTDHB equations 
in imaginary time and compute for static (time-independent) traps self-consistent
ground and excited eigenstates of bosonic systems, since in that case MCTDHB boils
down to the general variational many-body theory for interacting bosons (MCHB)
developed recently in \cite{MCHB_paper};
see \cite{Dieter_review} for the corresponding MCTDH case.
Second, with an essentially-exact and full 
many-body theory for the dynamics of bosonic systems 
we can now investigate strategies
for approximate many-body time-dependent theories,
e.g., when not all coefficients $\{C_{\vec{n}}(t)\}$ 
are employed in the ansatz for $\Psi(t)$.

\section{Numerical implementation}\label{secIII}

In the present section we describe the numerical 
implementation the developed MCTDHB theory for systems of cold bosonic atoms.
Here we would like to mention that in our independent technical implementation of MCTDHB 
we have borrowed much of the ideology of the implementation of the 
general MCTDH theory \cite{CPL,JCP,CMF,PR}.
Still, we reiterate two major differences,
being the direct utilization of bosons' statistics and
exploitation of two-body interactions which translate to
the appearance of the two-body density in the MCTDHB theory.

To integrate the time-dependent many-boson Schr\"odinger equation means
to find the many-body wavefunction at time $t$ by specifying the initial conditions, 
i.e., the many-body wavefunction at time zero.
In the framework of the MCTDHB theory we have to solve simultaneously 
Eqs.~(\ref{MCTDHB1_equ}) and (\ref{variation2_C_matrix}) by
specifying the initial set of expansion coefficients $\C(0)$ 
and respective orbitals $\bphi(0)$ at $t=0$.
Here and hereafter we use the column vector $\bphi=\{\phi_k\}$ 
to group together the orbitals.
In reality, the choice of the initial guess is quite a complicated task which depends on
a specific experimental setup and/or on the 
experimental sequence applied to the initial atomic cloud.
Typically, as an initial guess one uses a many-body solution 
of the stationary many-boson Schr\"odinger equation.

According to the MCTDHB derivation, 
Eq.~(\ref{variation2_C_matrix}) determines the evolution of the expansion 
coefficients for a given set of the orbitals and
Eq.~(\ref{MCTDHB1_equ}) governs the evolution of the 
orbitals for a given set of the expansion coefficients.
However, it is important to note that Eq.~(\ref{variation2_C_matrix}) determining 
the evolution of the expansion coefficients
does not depend explicitly on the orbitals, 
rather it depends on the one- and two-body matrix elements 
$h_{kq}$ and $W_{ksql}$, see (\ref{matrix_elements}).
Analogously, the expansion coefficients enter 
Eq.~(\ref{MCTDHB1_equ}) only implicitly 
via the elements of the one- and two-body densities 
$\rho_{kq}$ and $\rho_{ksql}$, 
see (\ref{DNS1}) and (\ref{DNS2}).
Let us display the MCTDHB equations (\ref{MCTDHB1_equ}) and (\ref{variation2_C_matrix})
in a form where these functional dependencies are explicitly indicated:
\beqn\label{OP}
& & i \dot{\C}(t) =
\O_1\!\left\{h_{kq}\left[\bphi(t)\right], W_{ksql}\left[\bphi(t)\right], \C(t)\right\} \nonumber \\
& & i \dot{\bphi}(t) =
\O_2\!\left\{\rho_{kq}\left[\C(t)\right], \rho_{ksql}\left[\C(t)\right], \bphi(t)\right\}, \
\eeqn
where the quantities $\O_1$ and $\O_2$ represent 
column vectors with functional dependences as indicated.
Before prescribing how to solve this coupled system simultaneously, 
let us first analyze each of these equations.
The first of these equations is linear (see Eq.~(\ref{variation2_C_matrix})) 
with respect to expansion coefficients $\C$ and thus,
if the matrix elements $h_{kq}$ and $W_{ksql}$ are given, 
can be effectively solved by integrators 
explicitly designed for linear equations, 
such as short iterative Lanczos (SIL) integrator \cite{SIL}.
The second equation for the orbitals is more sophisticated, 
because apart of the differential $\hat T$ and local $\hat V$ and $\hat W_{sl}$ 
(see Eq.~(\ref{one_body_pot})) operators,
it also contains the projection $ \hat {\mathbf P}$
classifying it as an integro-differential equation.
Again, if the elements of the 
one- and two-body densities $\rho_{kq}$ and $\rho_{ksql}$ are given,
this non-linear integro-differential equation can be propagated by 
means of general variable-order integrators, 
such as Adams-Bashforth-Moulton (ABM) predictor-corrector integrator \cite{ABM}
which is our choice here.
Now we are ready to integrate simultaneously the coupled system (\ref{OP}).
We recall that these equations are coupled through the time-dependent quantities 
$h_{kq}(t)$ and $W_{ksql}(t)$ which depend only on the orbitals, 
and the quantities $\rho_{kq}(t)$ and $\rho_{ksql}(t)$ 
which depend only on the expansion coefficients.
Therefore, if $t$ is discretized 
these numbers can be kept unchanged during each time-step.
Going along the MCTDH lines we apply a second 
order integration scheme with estimation on discretization 
error and adjustable time-step size $\tau$ \cite{CMF,PR}.
To propagate the coupled system (\ref{OP}) from 0 to $\tau$ we have 
used the following flowchart:
\begin{itemize}
\item Having at hand the initial conditions, i.e., 
the set of orbitals $\bphi(0)$ and expansion coefficients $\C(0)$ at $t=0$,
one computes the quantities $h_{kq}(0)$, $W_{ksql}(0)$ and $\rho_{kq}(0)$, $\rho_{ksql}(0)$, respectively.
\item Propagate $\C(0) \to \C(\tau/2)$ with SIL using $h_{kq}(0)$ and $W_{ksql}(0)$. \\
      Evaluate $\rho_{kq}(\tau/2)$ and $\rho_{ksql}(\tau/2)$ using $\C(\tau/2)$.
\item Propagate $\bphi(0) \to \bphi(\tau/2)$ with ABM using $\rho_{kq}(\tau/2)$ and $\rho_{ksql}(\tau/2)$.
\item For error estimation; Propagate 
      $\bphi(0) \to \widetilde{\bphi}(\tau/2)$ with ABM using $\rho_{kq}(0)$ and $\rho_{ksql}(0)$.
\item Propagate $\bphi(\tau/2) \to \bphi(\tau)$ with ABM using $\rho_{kq}(\tau/2)$, 
      $\rho_{ksql}(\tau/2)$.\\
      Evaluate $h_{kq}(\tau)$ and $W_{ksql}(\tau)$ using $\bphi(\tau)$.
\item Propagate $\C(\tau/2) \to \C(\tau)$ with SIL using $h_{kq}(\tau)$ and $W_{ksql}(\tau)$.
\item For error estimation; Backward propagate 
      $\widetilde{\C}(0) \gets \C(\tau/2)$ using $h_{kq}(\tau)$ and $W_{ksql}(\tau)$.
\end{itemize}
The differences $\C(0)-\widetilde{\C}(0)$ 
and $\bphi(\tau/2)-\widetilde{\bphi}(\tau/2)$ are used
to estimate the discretization error and to 
adjust the time-step size $\tau$.

We have so far fully implemented MCTDHB($M=2$), i.e., with two orbitals, in one-dimension,
and for a general two-body interparticle interaction $W(x-x')$
which is conveniently represented in a product form, $W(x-x')=\sum_lu_l(x)u_l(x')$.
In the context of quantum gases we have also implemented 
the popular contact interparticle interaction $W(x-x')=\lambda_0\delta(x-x')$, 
see Refs.~\cite{Leggett_review,Pethich_book,Stringari_book}. 
To represent the one-body Hamiltonian $\hat h(x)=\hat T(x)+V(x)$ and the orbitals 
we employ the discrete variable representation (DVR) technique \cite{DVR} 
based on harmonic oscillator, sinusoidal or exponential functions.
In the DVR basis the kinetic-energy differential operator $\hat T(x)$ corresponds to a
non-diagonal matrix, 
the potential $V(x)$ is diagonal, and the orbitals are represented by column vectors
on the respective DVR grid. 
Therefore, in the DVR technique a differentiation of the 
wavefunction is equivalent
to matrix to vector multiplication, 
whereas its integration 
corresponds to summation of the elements of the respective vector(s) with 
the corresponding DVR weights \cite{DVR}.
Finally, the accuracy of computations and respective integration efficiency 
of the numerical schemes described above depend on the specific many-boson
problem under investigation, 
see for specific examples the following section \ref{secIV}.

\section{Illustrative numerical examples}\label{secIV}

The first application of MCTDHB is for the dynamics of 
splitting a condensate when ramping-up a barrier such that a double-well is formed
and can be found elsewhere \cite{ramp_up_Letter}.
Before turning to the examples of this work
we first briefly summarize the main outcome of our application in \cite{ramp_up_Letter}.
We have found that the dynamics of splitting when ramping-up a barrier
depends on the duration of the process and on the (effective) interaction strength between the bosons.
There are (at least) two distinct regimes:
(i) an {\it adiabatic} regime where the initial condensed ground-state evolves towards
the ground two-fold fragmented eigenstate of the final double-well potential and
asymptotically approaches it with increasing rumping-up time;
(ii) an {\it inverse} regime where the initial condensed state
evolves towards the ground two-fold fragmented eigenstate only for short rumping-up times,
while for slow rumping-up processes the time-dependent state stays condensed during all the evolution
and thereby evolves to a {\it non}-ground many-body eigenstate, see \cite{ramp_up_Letter} for details.

The main purpose of our studies performed here 
is to compare the mean-field dynamics calculated by the 
time-dependent Gross-Pitaevskii equation and the many-body dynamics calculated with MCTDHB,
and thereby study some differences between 
the corresponding dynamics in simple and deductive examples.

In the numerical examples below we consider a many-boson system in one dimension.
We choose for convenience a length scale $L$ such that the energy unit is $\frac{\hbar^2}{L^2m}=1$, 
where $m$ is the mass of a boson, 
and then translate the time-dependent many-boson Schr\"odinger 
equation to dimensionless units.
The one-body Hamiltonian then reads 
$\hat h(x)=-\frac{1}{2}\frac{\partial^2}{\partial x^2} + V(x)$.
Of course we have also to choose a specific shape for the interparticle interaction
and we do so by taking the popular contact interaction 
$W(x-x')=\lambda_0\delta(x-x')$, see Refs.~\cite{Leggett_review,Pethich_book,Stringari_book}. 
The resulting two-body matrix elements (\ref{matrix_elements}) 
and time-dependent potentials (\ref{one_body_pot}) simplify,  
\beqn\label{W_delta}
 & & W_{ksql}(t) = \lambda_0 \int \phi_k^\ast(x,t) \phi_s^\ast(x,t)\phi_q(x,t) \phi_l(x,t) dx, \qquad
     W_{ks[ql]} = 2W_{ksql}, \nonumber \\
 & & \hat W_{sl}(x,t) = \lambda_0 \phi_s^\ast(x,t)\phi_l(x,t). \
\eeqn
With these quantities computed, 
Eq.~(\ref{MCTDHB1_equ}) for the evolution of the orbitals reads,
\beq
i\left|\dot\phi_j\right> = \hat {\mathbf P} \left[\hat h \left|\phi_j\right> + \lambda_0 \sum^M_{k,s,q,l=1} 
  \left\{\brho\right\}^{-1}_{jk} \rho_{ksql} \phi_s^\ast\phi_l 
\left|\phi_q\right> \right], \qquad 
\hat {\mathbf P} = 1-\sum_{j'=1}^{M}\left|\phi_{j'}\left>\right<\phi_{j'}\right|, 
\eeq
and the matrix elements of the Hamiltonian 
$H_{\vec{n}\vec{n}'}(t) = \left<\vec{n};t\left|\hat H\right|\vec{n}';t\right>$
needed for the propagation of the coefficients 
in (\ref{variation2_C_matrix}) are readily evaluated.

As illustrative numerical examples we consider the following scenario.
We prepare a system comprising of $N$ bosons 
in the ground-state $\Psi(0)$ 
of the double-well potential $V_0(x)=\frac{x^2}{2\sigma^2} + 8 e^{-x^2/(2\sigma^2)}, \ \sigma=2.6$.
The initial state $\Psi(0)$ is computed by imaginary time propagation
of MCTDHB(2) and separately by the Gross-Pitaevskii equation
for the sake of later comparison.
At time $t=0$ we halve the barrier between the two
wells and additionally translate the whole potential to the left.
The resulting double-well 
potential is $V(x)=\frac{(x+2)^2}{2\sigma^2} + 4 e^{-(x+2)^2/(2\sigma^2)}, \ \sigma=2.6$.
The many-boson state $\Psi(0)$ in which the system is prepared is 
obviously not in a stationary state any more 
and the interacting system is let to evolve in time.
The time-dependent many-boson wavefunction is respectively
computed by now real time propagating of MCTDHB(2) 
and of the Gross-Pitaevskii equation.
Two systems are considered.
The first with $N=100$ bosons and $\lambda_0=9.99/99=0.01009$,
the second with $N=1000$ bosons and $\lambda_0=9.99/999=0.0100$.
Note that the two systems are characterized by 
the same 'mean-field' factor $\lambda_0(N-1)=9.99$.
This means that {\it both} systems would show 
{\it identical} mean-field dynamics, since the only factor
concerning the number of bosons and their mutual interaction
entering the Gross-Pitaevskii theory 
is the product $\lambda_0(N-1)$ \cite{Leggett_review,Pethich_book,Stringari_book}.

Before moving to present and analyze the results, 
we would like to record some technical data used in our calculations. 
For the DVR we use 257 sinusoidal functions,
which we find to be sufficient to fully converge the present results.
The ABM integrator employed to propagate 
the orbitals (see section \ref{secIII}) is set to 7-$th$ order.
For $N=100$ bosons the maximal size of the SIL subspace needed for convergence
is as low as 8 vectors, whereas for $N=1000$ bosons it comprises of only 25 vectors!
Average values for the integration time-step size $\tau$
are $0.007$ and $0.002$ for $N=100$ and $N=1000$ bosons, respectively.
Finally, in all calculations and with the above parameters
the integration error of $\Psi(t)$ is $10^{-8}$,
combining the errors introduced by the SIL and ABM integrators
(see section \ref{secIII}).

\begin{figure}
\includegraphics[width=12cm,angle=0]{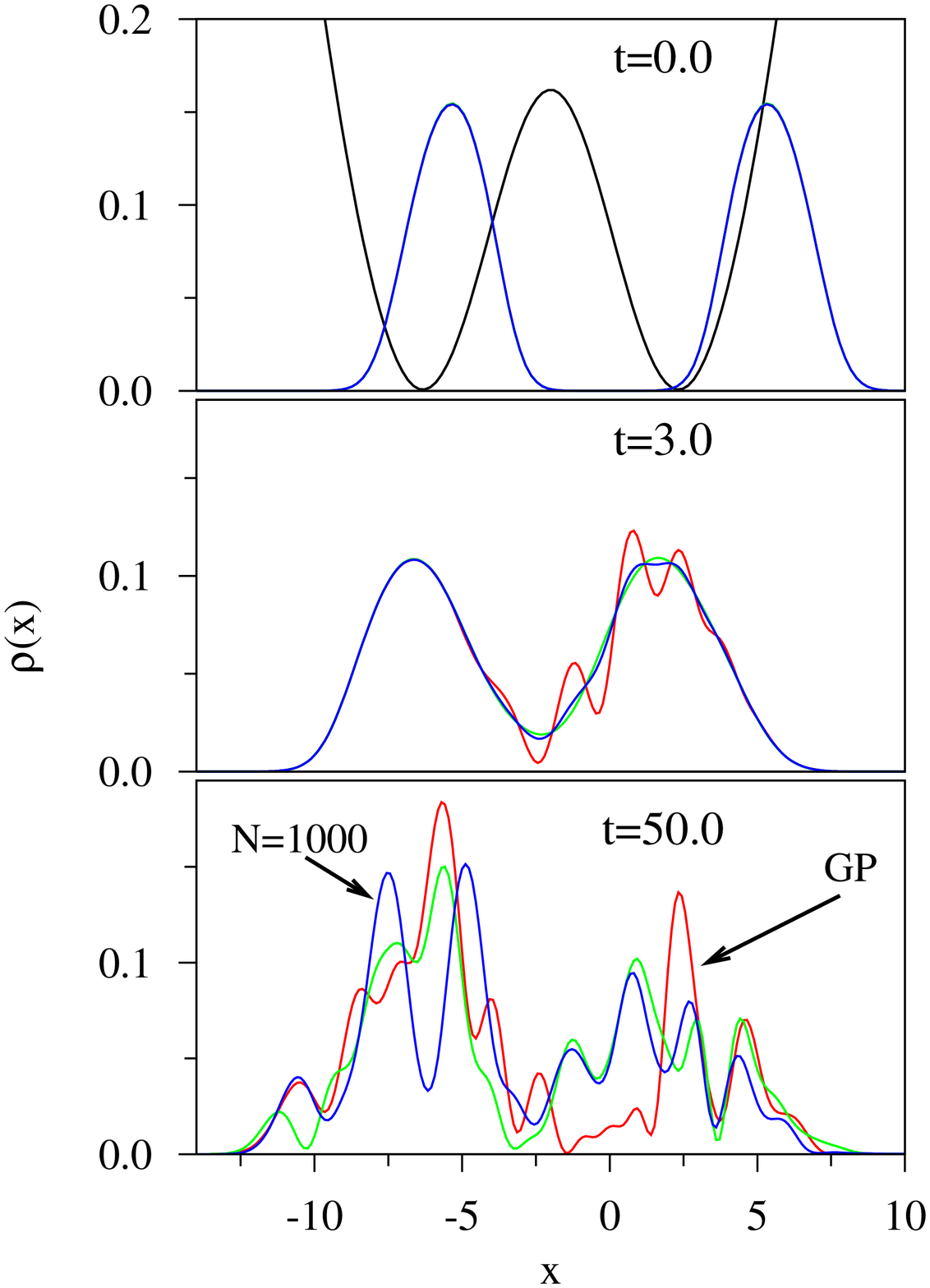}
\caption{(Color online) Many-body time evolution of two condensates 
made of $N=100$ (in green) and $N=1000$ (in blue) bosons 
in a double-well potential with corresponding interaction strengths 
$\lambda_0=\frac{9.99}{N-1}$ computed by the MCTDHB(2) theory.
For comparison, the corresponding solution computed by the time-dependent 
Gross-Pitaevskii equation is also displayed (in red).
Shown are three time snapshots of the density $\rho(x,t)$, normalized to 1.
At $t=0$ all densities are identical on the scale show.
To guide the eye, the double-well trap potential 
in which the condensates evolve
is also illustrated (in black).
The quantities shown are dimensionless.
See text for more details. 
}
\label{fig1}
\end{figure}

\begin{figure}
\includegraphics[width=11cm,angle=-90]{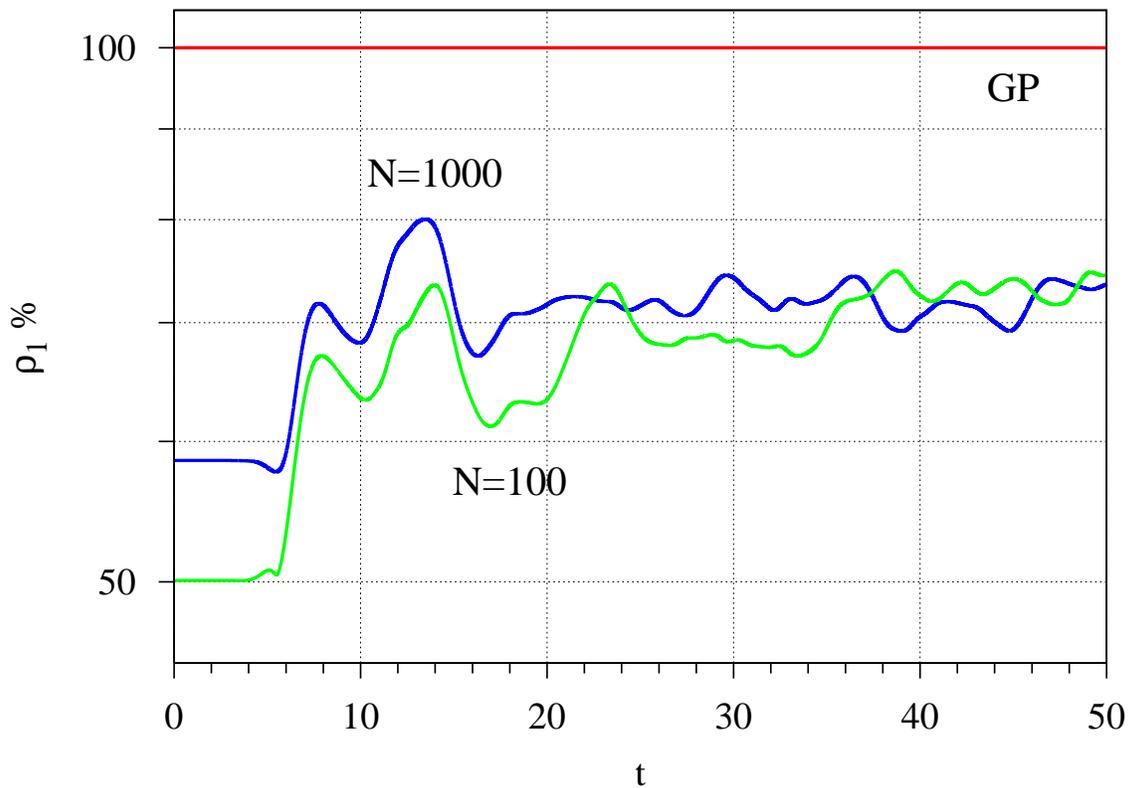}
\vglue 1.0 truecm
\caption{(Color online) Natural occupation numbers
of two condensates made of $N=100$ (in green) and $N=1000$ (in blue) bosons in a double-well potential
(same parameters as in Fig.~\ref{fig1}).
Shown in percents is the largest occupation $\rho_1$ ($\rho_2=100\%-\rho_1$; $\rho_2$ is not shown).
As time passes, the natural occupation numbers change substantially.
For comparison, the corresponding fixed $100\%$ occupation 
of the time-dependent Gross-Pitaevskii theory is also displayed (in red).
The quantities shown are dimensionless.
See text for more details. 
}
\label{fig2}
\end{figure}

We compare the time-dependent dynamics of the mean-field 
Gross-Pitaevskii and of the many-body MCTDHB(2) approach.
Since visualization of the time-dependent many-body wavefunction $\Psi(t)$ is quite cumbersome,
we plot in Fig.~\ref{fig1} snapshot of the density, 
$\rho(x,t)=\sum^2_{k,q=1} \rho_{kq}(t)\phi^\ast_k(x,t)\phi_q(x,t)$,
at different times
and compare with the respective Gross-Pitaevskii theory.
Furthermore, in Fig.~\ref{fig2} we display the natural occupation numbers $\rho_j(t),\, j\!=\!1,2$ of $\Psi(t)$,
i.e., the eigenvalues of the corresponding reduced one-particle density,
$\rho(x|x';t)=\sum^2_{j=1} \rho_j(t)\phi^{\ast{NO}}_j(x',t)\phi^{NO}_j(x,t)$,
at each point in time.
Of course, in Gross-Pitaevskii theory there
is only one occupation number, $\rho_1=100\%$.

Let us analyze the densities shown in Fig.~\ref{fig1}.
Since the barrier between the two potential wells is initially quite high, 
the corresponding three densities coincide at $t=0$, see top panel of Fig.~\ref{fig1}.
At $t=3.0$ we already see an observable difference.
The Gross-Pitaevskii density acquires wiggles due to collisions with the potential walls.
On the other hand, the densities in the many-body calculations
show almost no wiggles for $N=100$ and $N=1000$ bosons.
By $t=50.0$ all three densities have substantially distorted from
the initial conditions and exhibit multiple density wiggles of various sizes and depths. 
All densities are different from one another.
In particular, the systems with $N=100$ and $N=1000$ bosons which,
as we have mentioned above, 
have the {\it same} time evolution on the mean-field level
are actually showing quite {\it distinct} 
evolutions on the many-body level, see Fig.~\ref{fig1}.

Next, let us analyze the natural occupation numbers of $\Psi(t)$.
In the Gross-Pitaevskii dynamics the initial condition,
which is the ground state of the potential $V_0(x)$ computed by 
imaginary-time propagation of the Gross-Pitaevskii equation,
is fully coherent and hence there is only one occupation number, $\rho_1=100\%$.
This property of the condensate obviously does not change
in time when the time-dependent Gross-Pitaevskii equation governs the dynamics.
Contrastly, 
on the many-body level the situation differs substantially.
First, the initial conditions for $N=100$ and $N=1000$ bosons,
which are the respective ground states of the potential $V_0(x)$ computed by 
imaginary-time propagation of the MCTDHB(2) equations, 
actually correspond to fragmented condensates, 
see Fig.~\ref{fig2} at $t=0$.
Most importantly,
in the course of the many-body time evolution of the bosonic systems 
the natural-orbital occupations change in time by at least $20\%$.
This suggests an intricate and important property of the many-body dynamics which
by definition cannot be accessed by the mean-field dynamics. 

All in one, the numerical results of the application of MCTDHB(2) to
simple trapped bosonic systems demonstrate that the many-boson dynamics in traps 
is manageable and intriguing, 
and that it can also be quite different 
from the mean-field dynamics. 

\section{Summary and conclusions}\label{secV}

The evolution of Bose-Einstein condensates has been extensively studied
by the well-known time-dependent Gross-Pitaevskii equation
which assumes all bosons to reside in the same time-dependent orbital
and hence the condensate to be coherent at all times.
In this work we address the evolution of condensates on the many-body level,
and develop and report on an essentially-exact and
numerically-efficient approach for the solution of
the time-dependent many-boson Schr\"odinger equation.
We term our approach multi-configurational time-dependent Hartree for bosons (MCTDHB).

In the MCTDHB, the ansatz for the many-boson wavefunction 
is taken as a linear combination of all possible {\it time-dependent} permanents
made by distributing $N$ bosons over $M$ orthogonal {\it time-dependent} orbitals. 
The evolution of the many-body wavefunction 
is then determined by utilizing a standard time-dependent
variational principle -- either the Lagrangian formulation or the Dirac-Frenkel variational principle.
Performing the variation,
we arrive at two sets of coupled equations-of-motion,
one for the orbitals and one for the expansion coefficients. 
The first set comprises of first-order differential equations in time
and non-linear integro-differential equations in position space.
The second set consists of first-order differential equations  
with time-dependent coefficients.
MCTDHB naturally relates -- by performing propagation in imaginary time -- 
to the recently developed and successfully employed 
general variational mean-body theory with complete self-consistency
for {\it stationary} many-boson systems (MCHB).
From another instructive perspective,
MCTDHB can be seen as specification of the successful 
multi-dimensional wavepacket-propagation approach MCTDH to systems of identical bosons.
By explicitly exploiting bosons' statistics 
and the fact that only a two-body interaction is needed
it is now possible to successfully and quantitatively attack the dynamics 
of a much larger number of bosons with the 
developed MCTDHB theory.

With an essentially-exact and full many-body theory for the dynamics of bosonic systems, 
we can now investigate the many-body correlation effects coming atop the 
time-dependent Gross-Pitaevskii \cite{GPO1,GPO2}
and multi-orbital \cite{OAL_PLA_2007} mean-field 
dynamics of unfragmented and fragmented Bose-Einstein condensates, respectively.
Another promising direction would be 
to utilize MCTDHB for devising and investigating 
strategies for approximate many-body time-dependent approaches
such as Hilbert-space truncation schemes.

Finally, illustrative numerical examples of 
the many-body evolution of condensates in a one-dimensional 
double-well trap are provided, 
demonstrating an intricate dynamics of the density and natural orbitals as time progresses
which are not accounted for by the respective
Gross-Pitaevskii mean-field dynamics. 
This is the tip of the iceberg of many-body dynamical properties 
of Bose-Einstein condensates and interacting bosons
which await many more investigations with MCTDHB.

\begin{acknowledgments}
We thank Hans-Dieter Meyer for many helpful discussions and 
for making available the numerical integrators.
Financial support by the Deutsche Forschungsgemeinschaft is gratefully acknowledged.
\end{acknowledgments}

\appendix
\section{Matrix elements of two-body operators and of the two-body density matrix}\label{A1}

Matrix elements of the two-body operator $\hat W$
between two general permanents $\left|\vec{n};t\right>$ and $\left|\vec{n}';t\right>$.
The permanent denoted by $\left|\vec{n}_q^k;t\right>$ differs from $\left|\vec{n};t\right>$
by an excitation of one boson from the $q$-th to the $k$-th orbital,
whereas $\left|\vec{n}_{ql}^{ks};t\right>$ differs from $\left|\vec{n};t\right>$
by excitations of two bosons, one boson from the $q$-th to the $k$-th orbital
and a second boson from the $l$-th to the $s$-th orbital:
\beqn\label{matrix_elements_2b}
& & \left<\vec{n};t\left|\hat W\right|\vec{n};t\right> = 
\frac{1}{2}\sum_{j=1}^M n_j \left[\left(n_j -1\right) W_{jjjj}
+ \sum_{\{i\ne j\}=1}^M n_i W_{ji[ji]}\right], \nonumber \\
& & \left<\vec{n}_q^k;t\left|\hat W\right|\vec{n};t\right> =
 \sqrt{(n_k+1)n_q} \left[n_kW_{kkkq} + (n_q-1)W_{kqqq} + \sum_{\{i\ne q,k\}=1}^M 
n_i W_{ki[iq]}\right] \nonumber \\
& & \left<\vec{n}_{qq}^{kk};t\left|\hat W\right|\vec{n};t\right> =
 \frac{1}{2} \sqrt{(n_k+1)(n_k+2)(n_q-1)n_q} W_{kkqq} \nonumber \\
& & \left<\vec{n}_{ql}^{kk};t\left|\hat W\right|\vec{n};t\right> =
  \sqrt{(n_k+1)(n_k+2)n_qn_l} W_{kkql} \nonumber \\
& & \left<\vec{n}_{qq}^{ks};t\left|\hat W\right|\vec{n};t\right> =
  \sqrt{(n_k+1)(n_s+1)(n_q-1)n_q} W_{ksqq} \nonumber \\
& & \left<\vec{n}_{ql}^{ks};t\left|\hat W\right|\vec{n};t\right> =
  \sqrt{(n_k+1)(n_s+1)n_qn_l} W_{ks[ql]}.
\eeqn
Different indices $k,s,q,l$ cannot have the same value. 
All other matrix elements vanish. 

Matrix elements of the reduced two-body density 
$\rho(\r_1,\r_2|\r'_1,\r'_2;t)=\left<\Psi(t)\left|\hat{\mathbf \Psi}^\dag(\r'_1)\hat{\mathbf \Psi}^\dag(\r'_2)
\hat{\mathbf \Psi}(\r_1)\hat{\mathbf \Psi}(\r_2)\right|\Psi(t)\right>$
of the many-boson state $\left|\Psi(t)\right>=\sum_{\vec{n}}C_{\vec{n}}(t)\left|\vec{n};t\right>$:
\beqn
 & & \rho_{kkkk} = \sum_{\vec n} C_{\vec n}^\ast C_{\vec n} \left(n^2_k-n_k\right),  \nonumber \\
 & & \rho_{ksks} = \sum_{\vec n} C_{\vec n}^\ast C_{\vec n} n_kn_s, \nonumber \\
 & & \rho_{kkqq} = \sum_{\vec n} C_{\vec n}^\ast C_{\vec n_{kk}^{qq}} \sqrt{(n_k-1)n_k(n_q+1)(n_q+2)},
\nonumber \\
& & \rho_{kkkl} = \sum_{\vec n} C_{\vec n}^\ast C_{\vec n_{k}^{l}} (n_k-1)\sqrt{n_k(n_l+1)}, \nonumber \\
& & \rho_{ksss} = \sum_{\vec n} C_{\vec n}^\ast C_{\vec n_{k}^{s}} n_s\sqrt{n_k(n_s+1)}, \nonumber \\
& & \rho_{kkql} = 
\sum_{\vec n} C_{\vec n}^\ast C_{\vec n_{kk}^{ql}} \sqrt{(n_k-1)n_k(n_q+1)(n_l+1)}, \nonumber \\
& & \rho_{ksqq} = 
\sum_{\vec n} C_{\vec n}^\ast C_{\vec n_{ks}^{qq}} \sqrt{n_kn_s(n_q+1)(n_q+2)}, \nonumber \\
& & \rho_{kssl} = 
\sum_{\vec n} C_{\vec n}^\ast C_{\vec n_{k}^{l}} n_s\sqrt{n_k(n_l+1)}, \nonumber \\
& & \rho_{ksql} = 
\sum_{\vec n} C_{\vec n}^\ast C_{\vec n_{ks}^{ql}} \sqrt{n_kn_s(n_q+1)(n_l+1)}. \
\eeqn
Different indices $k,s,q,l$ cannot have the same value. 
All other matrix elements can be computed due to the symmetries of the two-body operator,
$\rho_{ksql}=\rho_{kslq}=\rho_{sklq}=\rho_{skql}$,
and its hermiticity, $\rho^\ast_{ksql}=\rho_{qlks}$.

\section{Derivation of the MCTDHB equations-of-motion starting from 
the Dirac-Frenkel variational principle}\label{A2}

The Dirac-Frenekl variational principle reads \cite{DF1,DF2}:
\beq\label{DF_VP}
 \left<\delta\Psi(t)\left|\hat H -i\frac{\partial}{\partial t}\right|\Psi(t)\right> = 0.
\eeq
Given the MCTDHB ansatz for the many-boson wavefunction, 
$\left|\Psi(t)\right> = \sum_{\vec{n}}C_{\vec{n}}(t)\left|\vec{n};t\right>$, 
the variation $\delta\Psi(t)$ in Eq.~(\ref{DF_VP}) is performed separately
with respect to the expansion coefficients $\{C_{\vec{n}}(t)\}$
and with respect to the orbitals $\{\phi_k(\r,t)\}$.

We begin as in the main text with the coefficients. 
From the structure of $\left|\Psi(t)\right>$ as a sum of permanents one obviously has
\beq\label{variation_psi_c}
  \frac{\partial\left<\Psi(t)\right|}{\partial C^\ast_{\vec{n}}(t)} = \left<\vec{n};t\right|.
\eeq
Hence, making use of the orthonormality relation of the permanents 
$\left<\vec{n};t\left|\right.\vec{n}';t\right>=\delta_{\vec{n},\vec{n}'}$
one straightforwardly finds:
\beq\label{DF_inter0}
  \left(\frac{\partial}{\partial C^\ast_{\vec{n}}(t)}
    \left<\Psi(t)\left|\left)\hat H -i\frac{\partial}{\partial t}\right|\Psi(t)\right>\right.\right.
  = 0 \ \ \Longrightarrow
  \bcalH(t) \C(t) = i\frac{\partial \C(t)}{\partial t},
\eeq
which is precisely the result depicted in Eq.~(\ref{variation1_C_matrix}).

To proceed for the orbitals, 
we need a few identities and relations.
Starting from the basic definitions 
$\hat{\mathbf \Psi}(\r)=\sum_k b_k(t)\phi_k(\r,t)$ and 
$b_k(t) = \int \phi_k^\ast(\r,t) \hat{\mathbf \Psi}(\r) d\r$
connecting the annihilation and field operators one readily has,
\beq\label{variation_b}
 \frac{\delta b_k(t)}{\delta\phi_q^\ast(\r,t)} = \delta_{kq}\hat{\mathbf \Psi}(\r).
\eeq
With the help of (\ref{variation_b}) we can take the variation of a single permanent
$\left<\vec{n};t\right|$,
\beq\label{single_permanent}
   \frac{\delta\left<\vec{n};t\right|}{\delta\phi_k^\ast(\r,t)} = 
\left<\vec{n};t\right|b_k^\dag(t)\hat{\mathbf \Psi}(\r),
\eeq
 and from it of the 'bra' $\left<\Psi(t)\right|$ itself,
\beq\label{variation_psi_phi}
 \frac{\delta\left<\Psi(t)\right|}{\delta\phi_k^\ast(\r,t)} = 
 \left<\Psi(t)\right|b_k^\dag(t)\sum_q b_q(t)\phi_q(\r,t).
\eeq
On the 'ket' side of Eq.~(\ref{DF_VP}) we keep in mind that,
\beq\label{tricky_derivative}
 i\frac{\partial}{\partial t}\left|\Psi(t)\right> =\sum_{\vec{n}}
\left\{ i\frac{\partial C_{\vec{n}}(t)}{\partial t} + 
C_{\vec{n}}(t) i\frac{\partial}{\partial t} \right\} \left|\vec{n};t\right>. 
\eeq   
Utilizing Eqs.~(\ref{DF_inter0}), (\ref{variation_psi_phi}) and (\ref{tricky_derivative}) 
and the usual commutation relations between the bosonic annihilation and creation operators, 
we evaluate separately the contributions coming from the one-body 
$\hat h - i\frac{\partial}{\partial t}$ and two-body $\hat W$ operators.
Combining the results one finds: 
\beqn\label{DF_inter1}
& & \left(\frac{\delta}{\delta\phi^\ast_k(\r,t)}
    \left<\Psi(t)\left|\left)\hat H -i\frac{\partial}{\partial t}\right|\Psi(t)\right>\right.\right.
 = 0 \ \ \Longrightarrow \nonumber \\ \nonumber \\
& & \sum^M_{q=1}\left\{\rho_{kq} \left(\hat h - i\frac{\partial}{\partial t} \right) +
\sum^M_{s,l=1}\rho_{ksql} \hat{W}_{sl}\right\}\left|\phi_q\right> =
 -\sum^M_{j=1} 
\left<\Psi\left| b_k^\dag \left[\hat H - i\frac{\partial}{\partial t}, b_j\right]\right|\Psi\right>  
\left|\phi_j\right>. \
\eeqn
To show that (\ref{DF_inter1}) is exactly Eq.~(\ref{MCTDHB_mu}) 
we evaluate the commutator on the right-hand-side of the former,
\beq\label{ham_identity_DF}
 \left[\hat H - i\frac{\partial}{\partial t},b_j\right] =
 - \sum_{q}\left\{\left( h_{jq} - \left(i\frac{\partial}{\partial t}\right)_{jq} \right)b_q +
 \sum_{s,l} W_{jsql} b_s^\dag b_q b_l \right\}, 
\eeq
from which we readily get that the right-hand-side of (\ref{DF_inter1}) 
is nothing but (sum over) $\mu_{kj}(t)$ of Eq.~(\ref{MCTDHB_mu}), i.e., that
\beq\label{mu_DF}
 \mu_{kj}(t) = - 
\left<\Psi\left| b_k^\dag \left[\hat H - i\frac{\partial}{\partial t}, b_j\right] \right|\Psi\right>. 
\eeq
From this point on the two derivations coincide,
leading as expected to identical equations-of-motion of MCTDHB.

We conclude the appendix by pointing out an additional connection between the two variational 
formulations used in this work to derive the MCTDHB equations-of-motion.
Eqs.~(\ref{DF_inter1}), (\ref{ham_identity_DF}) and (\ref{mu_DF}) hint towards 
another role played by the Lagrange multipliers $\mu_{kj}(t)$ employed in the main text.
Since in the Dirac-Frenkel formulation the variation is taken before matrix elements are computed,
whereas in the Lagrangian formulation the situation reverses,
there would have been 'missing' terms in the equations-of-motion
derived by the Lagrangian formulation. 
The introduction of the Lagrange multipliers in Eq.~(\ref{action_functional})
exactly 'compensates' for these terms.

\end{document}